\documentclass[%
 reprint,
 superscriptaddress,
 amsmath,amssymb,
 aps,
prb,
]{revtex4-2}

\usepackage{graphicx}
\usepackage{dcolumn}
\usepackage{bm}
\usepackage{hyperref}

\usepackage{siunitx}
\usepackage{xcolor}

\begin{document}


\title{Plaid-Like Spin Splitting and Chirality of Magnon Bands in Antiferromagnetic MnTe$_2$}

\author{Dirk Wulferding}
\email{d.wulferding@sejong.ac.kr}
\affiliation{Department of Physics and Astronomy, Sejong University, Seoul 05006, Republic of Korea}

\author{Daehyeon An}
\affiliation{Department of Physics, Korea Advanced Institute of Science and Technology (KAIST), Daejeon 34141, Korea}

\author{Jiwon Choi}
\affiliation{Department of Physics, Korea Advanced Institute of Science and Technology (KAIST), Daejeon 34141, Korea}

\author{Dongmin Mun}
\affiliation{Department of Physics, Sungkyunkwan University, Suwon 16419, Republic of Korea}

\author{Youngsu Choi}
\affiliation{Department of Physics, Sungkyunkwan University, Suwon 16419, Republic of Korea}

\author{Sivasakthi Kuppusamy}
\affiliation{Institute of Physics, Academia Sinica, Nankang, Taipei 11529, Taiwan}

\author{Sritharan Krishnamoorthi}
\affiliation{Institute of Physics, Academia Sinica, Nankang, Taipei 11529, Taiwan}

\author{Raman Sankar}
\email{sankarndf@gmail.com}
\affiliation{Institute of Physics, Academia Sinica, Nankang, Taipei 11529, Taiwan}

\author{Myung Joon Han}
\email{mj.han@kaist.ac.kr}
\affiliation{Department of Physics, Korea Advanced Institute of Science and Technology (KAIST), Daejeon 34141, Korea}

\author{Se Kwon Kim}
\email{sekwonkim@kaist.ac.kr}
\affiliation{Department of Physics, Korea Advanced Institute of Science and Technology (KAIST), Daejeon 34141, Korea}

\author{Kwang-Yong Choi}
\email{choisky99@skku.edu}
\affiliation{Department of Physics, Sungkyunkwan University, Suwon 16419, Republic of Korea}

\date{\today}

\begin{abstract}
Altermagnets constitute an emerging class of magnetic materials that combine compensated antiferromagnetic order with spin-split excitations arising from crystalline symmetries. Despite strong theoretical interest, their experimental identification remains challenging. Here, we demonstrate that helicity- and angle-resolved Raman scattering measurements reveal reduced rotational symmetries of
magnons and a pronounced imbalance between left- and right-circular polarization channels, indicating momentum-dependent magnon handedness. First-principles DFT+$U$ calculations combined with linear spin-wave theory uncover a characteristic plaid-like spin-splitting structure in momentum space. The resulting magnon spin textures are dictated by the unconventional sublattice symmetries of $\text{MnTe}_2$ and closely emulate those of altermagnetic electronic bands. Our work provides evidence of chiral spin-wave excitations unique to this non-coplanar antiferromagnet.

\end{abstract}

\maketitle


\section{\label{intro}Introduction}
Altermagnets have recently emerged as a third fundamental class of magnetic materials, transcending the traditional dichotomy between ferromagnetism and antiferromagnetism. By uniquely bridging the features of both, altermagnets host a compensated collinear magnetic order while exhibiting large spin splitting in their electronic band structures~\cite{Smejkal-22a,Smejkal-22b,Hayami-19,Ahn-19,Naka-19,Hayami-20,Yuan-20,Liu-22,Mazin-22}.
Unlike conventional spintronic materials, this distinct behavior does not arise from relativistic spin–orbit coupling. Instead, it is a direct consequence of crystal symmetries that connect opposite magnetic sublattices via rotation or mirror operations rather than simple translations or inversion~\cite{Smejkal-22a,Smejkal-22b,Gonzalez-21,Hu-24,Krempasky-24}.

This concurrence of vanishing net magnetization and spin-polarized electronic states gives rise to a range of time-reversal-symmetry-breaking phenomena, including the giant anomalous Hall and spin Hall effects, non-vanishing Berry-curvature multipoles, and highly anisotropic spin-polarized currents, positioning altermagnets as a promising platform for next-generation spintronics~\cite{Ma-21,Smejkal-22c,Smejkal-22d,Hu-25,Song-25}. Beyond electronic transport, altermagnets are predicted to harbor chiral magnons.  Driven by a symmetry-odd, momentum-dependent exchange field, these excitations exhibit a nonreciprocal spin-wave dispersion,
$\omega(\mathbf{k})\neq \omega(-\mathbf{k})$, leading to directionally asymmetric magnon propagation~\cite{Smejkal-23,Maier-23}. While inelastic neutron scattering and angle-resolved photoemission spectroscopy experiments on the canonical hexagonal MnTe altermagnet have successfully evidenced such chiral splitting~\cite{lee-24,Liu-24}, experimental observations in other structural motifs remain sparse.

Recently, the conceptual framework of altermagnetism has expanded to encompass complex non-collinear magnetic systems. In this regime, non-trivial spin–momentum–locking textures emerge through anisotropic hopping and color symmetries that break parity–time-reversal ($\mathcal{PT}$) symmetry in reciprocal space~\cite{cheong-24a,cheong-24b,Radaelli-25,Rimmler-25}. A prominent candidate for this broader classification is MnTe$_2$, the three-dimensional cubic counterpart of the hexagonal MnTe altermagnet~\cite{Zhu-24}.

\begin{figure*}
    \centering
    \includegraphics[width=500pt]{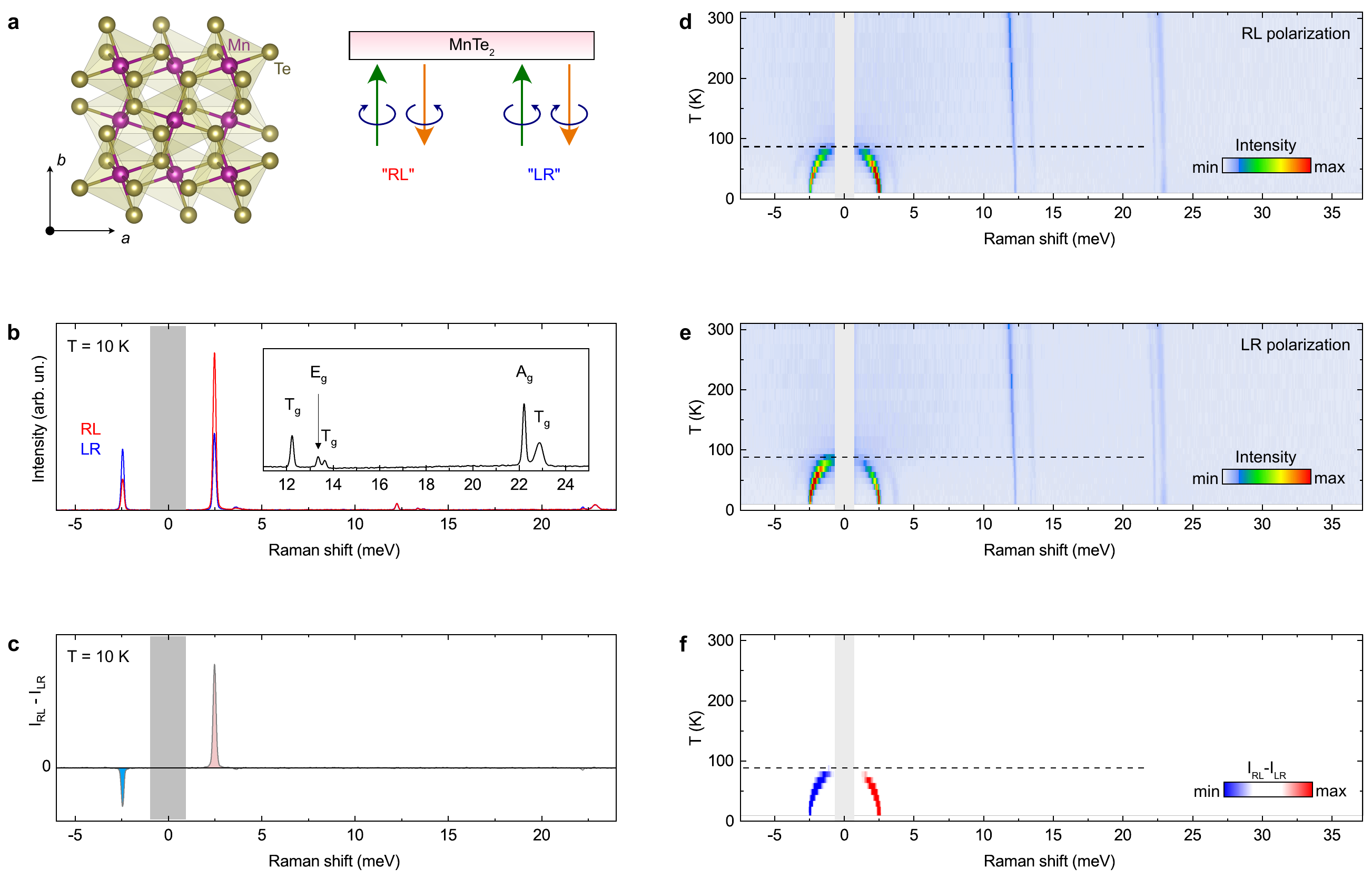}
    \caption{
    (a) Pyrite-type crystal structure of MnTe$_2$. The Raman scattering configurations $RL$ and $LR$ are schematically illustrated, where green and orange arrows indicate the incident monochromatic laser light and the inelastically scattered light, respectively. Circular arrows denote the helicity (circular polarization) of the photons. (b) Two representative spectra measured at $T = 10$ K in $RL$ (red) and $LR$ (blue) polarization. The inset presents a zoom-in of the optical phonons together with their symmetry assignments. (c) Intensity difference between the data taken in $RL$ and $LR$ shown in (b). (d,e) Color-contour plot of the temperature-dependent Raman spectra measured in $RL$ (d) and $LR$ (e) polarization configurations. (f) Color-contour map of the intensity difference ($I_{RL}-I_{LR}$) over the temperature range 10 K -- 305 K. The dashed black horizontal lines in panels (d), (e), and (f) mark $T_\mathrm{N}$. Gray areas around 0 meV in the color-contour plots mask the laser line.
    }
    \label{fig:Tdep}
\end{figure*}

MnTe$_2$ is a pyrite-type antiferromagnetic semiconductor (energy gap $E_g = 0.87$ eV), in which Mn$^{2+}$ ($S=5/2$) ions form a face-centered cubic network of corner-sharing MnTe$_6$ octahedra, as shown in Fig.~1(a)~\cite{structure}. Below the N\'{e}el transition temperature ($T_\mathrm{N} = 87$ K), the Mn moments across four sublattices orient along distinct local $[111]$ directions, resulting in a perfectly compensated unit cell~\cite{burlet-97,Chen-20}. Despite the theoretical classification of MnTe$_2$ as a non-collinear altermagnet, it remains an open question whether chiral magnetic excitations are detectable in such a complex multi-sublattice system.

In this paper, we address this question by investigating the magnetic excitations of MnTe$_2$ using helicity- and angle-resolved Raman spectroscopy, combined with density functional theory (DFT) + $U$ calculations and linear spin-wave theory. Our results reveal pronounced circular Raman dichroism and anomalous rotational symmetry in the magnetic excitation spectrum. These findings provide direct evidence for chiral magnons with a well-defined handedness, thereby unveiling the altermagnetic nature of the noncollinear magnet MnTe$_2$.

\section{\label{results}Results}
\subsection{Raman scattering}

The group-theoretical analysis for the $Pa\bar{3}$ space group with Mn occupying the Wyckoff positions Mn($4a$) and Te($8c$) yields a total of 6 Raman-active phonons: 1 A$_g$ + 1 $^1$E$_g$ + 1 $^2$E$_g$ + 3 T$_g$~\cite{aroyo-1, aroyo-2, aroyo-3}. Their corresponding Raman tensors are:

\begin{center}
\mbox{A$_{g}$=$\begin{pmatrix} a & 0 & 0\\ 0 & a & 0\\ 0 & 0 & a\\ \end{pmatrix}$,

$^1$E$_g$=$\begin{pmatrix} b+\sqrt{3}c & 0 & 0\\ 0 & b-\sqrt{3}c & 0\\ 0 & 0 & -2b\\ \end{pmatrix},$}
\end{center}

\begin{center}
\mbox{$^2$E$_g$=$\begin{pmatrix} c-\sqrt{3}b & 0 & 0\\ 0 & c+\sqrt{3}b & 0\\ 0 & 0 & 0\\ \end{pmatrix}$,}
\end{center}

\begin{center}
\mbox{T$_g$=$\begin{pmatrix} 0 & 0 & 0\\ 0 & 0 & d\\ 0 & d & 0\\ \end{pmatrix}$,

 T$_g$=$\begin{pmatrix} 0 & 0 & d\\ 0 & 0 & 0\\ d & 0 & 0\\ \end{pmatrix}$,

T$_g$=$\begin{pmatrix} 0 & d & 0\\ d & 0 & 0\\ 0 & 0 & 0\\ \end{pmatrix}$.}
\end{center}

The assignment of the observed phonons is detailed in the inset of Fig. 1(b), based on the phonons' characteristic polarization behavior. To characterize the magnetic excitations in MnTe$_2$, we first focus on their symmetry properties and thermal evolution. Figure 1(b) shows two representative Raman spectra measured in the cross-circular polarized $RL$ (red) and $LR$ (blue) configurations. The energy scale plotted extends into the anti-Stokes side for negative energy values. Note that these spectra were taken at $T=10$ K ($\approx 0.9$ meV). At such a low-temperature regime, anti-Stokes signals should normally vanish due to the thermal depletion of excited states.

Contrary to this expectation, a dominant excitation at 2.5 meV is seen on both the Stokes and anti-Stokes side, which, together with a weaker excitation at 3.6 meV, we attribute to magnon modes.
Notably, the lower-energy mode at 2.5 meV exhibits an anomalous intensity imbalance between the polarization  $RL$ and $LR$ channels, while simultaneously yielding an unexpectedly strong anti-Stokes response. Most strikingly, this $RL$/$LR$ imbalance inverts on the anti-Stokes side. Such anomalous behavior is characteristic of excitations with nonreciprocal or chiral character, where the breaking of time-reversal or combined spatial–temporal symmetries leads to directional or polarization-dependent scattering selection rules. We quantify this effect in Fig. 1(c) by plotting the differential intensity ($I_{RL}-I_{LR}$), which is proportional to the Raman circular dichroism (RCD). The resulting RCD spectrum demonstrates a pronounced chiral contrast for the 2.5 meV excitation, characterized by an opposite sign change between the Stokes and anti-Stokes sides.

Next, we trace the evolution of the polarization anomaly as a function of temperature to establish its direct connection to the magnetically ordered state. Figures 1(d),(e) present color-contour plots of the Raman scattering intensity in the temperature versus Raman shift measured in each $RL$ and $LR$ polarization.  Above the N\'{e}el temperature ($T > T_{\mathrm{N}}$), the phonon modes experience a gradual softening with increasing temperature, characteristic of conventional lattice vibrations, only with weak anharmonic renormalization. In this temperature regime, we also identify thermally damped magnetic excitations (paramagnons) extending up to about 10 meV, which set the overall magnon energy scale.
On the other hand, upon cooling below $ T_\mathrm{N}$, two highly intense modes evolve from quasi-elastic scattering and progressively harden, reaching energies of approximately 2.5 meV and 3.6 meV at base temperature. Their thermal evolution follows a characteristic order-parameter–like behavior.

Finally, the full temperature dependence of the RCD signal ($I_{RL}-I_{LR}$) is plotted in Fig.~1(f). For all temperatures below $T_\mathrm{N}$, the Stokes/Anti-Stokes intensity imbalance between the $RL$ and $LR$ channel is visible mainly for the lower-energy magnon branch at 2.5 meV.
Conversely,  neither the phonon modes nor the higher-energy magnon at 3.6 meV  exhibit any detectable chiral asymmetry, suggesting that the chiral anomaly is exclusive to the lower-energy magnon, providing a fingerprint of the altermagnetic phase in MnTe$_2$.

\begin{figure*}
    \centering
    \includegraphics[width=500pt]{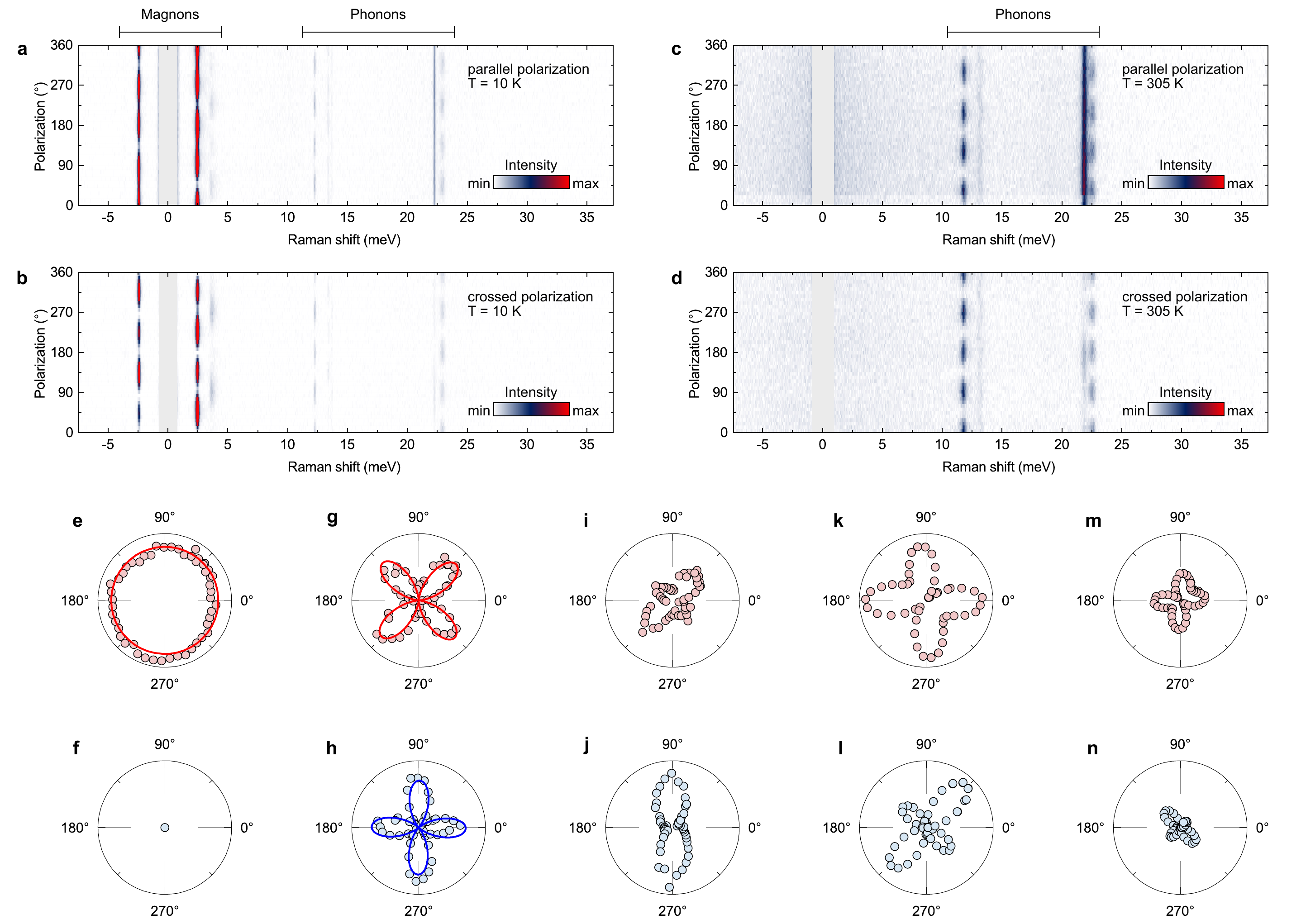}
    \caption{
    (a) and (b): Color-contour plots of polarization-resolved Raman data obtained at $T = 10$ K in parallel and crossed polarization, respectively. (c) and (d): Polarization-resolved Raman data collected at $T = 305$ K in parallel and crossed polarization, respectively. Gray areas around 0 meV mask the laser line. (e) and (f): Intensity profiles of the $A_g$ phonon at 22.2 meV measured at $T = 10$ K in parallel and crossed polarization-respectively. (g) and (h): Intensity profiles of the $T_g$ phonon at 12.2 meV measured at $T = 10$ K in parallel and crossed polarization-respectively. (i) and (j): Intensity profiles of the magnon branch at 3.6 meV measured in parallel and crossed polarization-respectively. (k) and (l): Intensity profiles of the magnon branch at 2.5 meV measured in parallel and crossed polarization-respectively. (m) and (n): Intensity profiles of the anti-Stokes component of the magnon branch at -2.5 meV measured in parallel and crossed polarization-respectively. Solid lines in panels (e)-(h) represent fits based on the Raman tensors.
    }
    \label{fig:Tdep}
\end{figure*}

A full characterization of the symmetries of Raman-active excitations hinges on tracing their Raman-intensity profiles while performing in-plane rotation of linearly polarized light, measured in both parallel ($\textbf{e}_{\mathrm{in}}$ // $\textbf{e}_{\mathrm{out}}$) and crossed ($\textbf{e}_{\mathrm{in}} \perp \textbf{e}_{\mathrm{out}}$) configurations. Through this procedure, we gain direct access to the individual elements of the Raman tensors. Figures 2(a)-(d) show the resulting angle-resolved polarization-dependent Raman intensity measured in the magnetically ordered state ($T = 10$ K; panels (a)-(b)) and in the paramagnetic phase above $T_\mathrm{N}$ ($T = 305$ K; panels (c)-(d)), for both polarization geometries. The phonons occupy the 10--25 meV range and display either isotropic or fourfold angular symmetries, which is consistent with the underlying Raman tensors for the pyrite structure. We also note that there is no change in the symmetry of phonon modes when entering the magnetically ordered phase.

Below 5 meV, the two magnon modes appear, with the lower-energy mode showing a strong contribution to the anti-Stokes response. Before investigating the magnetic excitations in detail, we first focus on two representative phonon modes. We plot the angular intensity profiles of the $A_g$ phonon located at 22.2 meV in parallel (Fig. 2(e)) and in crossed (Fig. 2(f)) polarization. As expected from its Raman tensor, the phonon intensity is isotropic in the parallel polarization, and vanishes in the crossed polarization. Likewise, the angular intensity profiles of the $T_g$ phonon at 12.2 meV show fourfold symmetry in both parallel (Fig. 2(g)) and crossed (Fig. 2(h)) polarization, with a relative phase shift of 45$^{\circ}$  between the two configurations, again fully consistent with its Raman tensor.

We now turn our attention to the magnetic excitations, starting with the branch at 3.6 meV. Although its overall intensity is relatively weak, we can still track its angular dependence, as summarized in the polar plots of Fig. 2(i),(j). In contrast to the phonon modes that strictly follow the symmetry-allowed Raman tensors, this excitation exhibits a pronounced twofold modulation in the parallel polarization, which becomes even more prominent in the crossed polarization. We recall that one-magnon scattering, which involves a spin change of $\Delta S = \pm 1$, requires the Raman tensor to possess nonzero off-diagonal antisymmetric components. In the $Pa\bar{3}$ space group, one-magnon excitations are therefore generally expected to appear in the $T_g$ (triply degenerate) symmetry channel. In an altermagnet, however, the $T_g$ magnons can acquire a defined handedness, resulting in different scattering intensities for the $RL$ and $LR$ circular polarization. This chirality-induced intensity imbalance gives rationale for the emergence of a twofold rotational symmetry, reduced from the fourfold symmetry.

Figures 2(k),(l) show the angular intensity profiles of the lower-lying 2.5 meV branch on the Stokes side in parallel and crossed polarizations, respectively. While the profile in Fig. 2(k) follows an almost fourfold behavior, there is a subtle distortion most clearly seen by the absence of fully-developed nodes around 135$^{\circ}$ and 315$^{\circ}$. More prominently, in the crossed configuration shown in Fig. 2(l), the fourfold symmetry is clearly convoluted with a twofold component, producing intensity maxima at 45$^{\circ}$ and 225$^{\circ}$. Finally, we compare the Stokes-side with the anti-Stokes response for the 2.5 meV magnon branch in panels (m) and (n). Strikingly, the polarization patterns appear rotated by 90$^{\circ}$ relative to each other, counter-acting the behavior observed on the Stokes-side. The 90$^{\circ}$ rotation between the $RL$ and $LR$ signal intensities may be associated with the $d$-wave symmetry of the altermagnetic order parameter in momentum space. Switching from $RL$ to $LR$ polarization effectively probes the opposite spin sublattice. Consequently, the intensity maxima occur at momentum-space locations where the $d$-wave order parameter has the opposite sign or phase, leading naturally to a 90$^{\circ}$ rotation of the intensity pattern between the two channels.

\subsection{DFT calculations}

To construct an effective spin Hamiltonian for MnTe$_{2}$, we performed first-principles density functional theory (DFT) and magnetic force linear response calculations. In this framework, we considered exchange interactions up to first- and second-nearest neighbors in the form of
\begin{equation}
\begin{aligned}
\mathcal{H}_{\mathrm{eff}}
= {} & \sum_{\langle i,j\rangle}
\Big(
J_1\,\vec{S}_i \cdot \vec{S}_j
+ \vec{D}_{1} \cdot (\vec{S}_i \times \vec{S}_j)
+ \vec{S}_i \mathbf{\Gamma}_{1}  \vec{S}_j
\Big) \\
& + \sum_{\langle\langle i,j\rangle\rangle}
\Big(
J_2\,\vec{S}_i \cdot \vec{S}_j
+ \vec{D}_{2} \cdot (\vec{S}_i \times \vec{S}_j)
+ \vec{S}_i  \mathbf{\Gamma}_{2}  \vec{S}_j
\Big),
\end{aligned}
\end{equation}
where $\vec{S}_{i,j}$ denotes the spin operators at Mn sites $i$ and $j$. ${\langle i,j \rangle}$ and ${\langle\langle i,j \rangle\rangle}$ refer to first- and second-nearest-neighbor Mn pairs, respectively. The isotropic exchange interaction $J_n$ is a constant, $\vec{D}_n$ corresponds to a Dzyaloshinskii-Moriya vector, and the anisotropic exchange $\mathbf{\Gamma}$ is given by a $3\times3$ symmetric tensor.

The calculated exchange parameters are summarized in Table~\ref{Table_J}. While the dominant interaction is the antiferromagnetic nearest-neighbor coupling $J_1 = 0.675$~meV, the second-nearest-neighbor exchange $J_2 = -0.029$~meV is ferromagnetic and much weaker, with a magnitude comparable to that of $\vec{D}_1$.  As expected from the presence of an inversion center at the midpoint of each second-neighbor Mn--Mn bond, the corresponding DM interaction vanishes, $\vec{D}_2 = 0$. By contrast,
the first-neighbor DM interaction, with magnitude $|\vec{D}_1| = 0.063$~meV plays an important role in stabilizing the noncollinear spin ordering. The inclusion of further-neighbor exchanges $J_{n=3, 4}$ just shifts the $\Gamma$-point magnon energies by $\sim$8\%. Overall, the two leading exchange interactions are comparable in both magnitude and sign to the values $J_1\approx 7.1$~K and  $J_2\approx -1.3$~K reported in previous work~\cite{Okada-77}.

\begin{table*}[htb!]
\centering
\caption{Calculated first- and second-neighbor magnetic interaction parameters in units of meV. Here, with Mn$_1$(0, 0, 0) as the reference, $n=1$ and $n=2$ correspond to its pairs with Mn$_2$(-0.5, 0.5, 0) and Mn$_1$(0, 0, 1), respectively (see also Table~\ref{Table1}). All other symmetry-equivalent interactions are obtained via the magnetic space group $Pa\bar3$.}
\setlength{\tabcolsep}{8pt}
\resizebox{\textwidth}{!}
{
\begin{ruledtabular}
\begin{tabular}{c|c|c|c|c|c|c|c|c|c|c}

& $J_n$
& $D_n^{x}$ & $D_n^{y}$ & $D_n^{z}$
& $\Gamma_n^{xx}$ & $\Gamma_n^{yy}$ & $\Gamma_n^{zz}$
& $\Gamma_n^{xy}$ & $\Gamma_n^{yz}$ & $\Gamma_n^{zx}$ \\ \hline

$n=1$
&    0.675
&    0.006 & $-$0.026 &    0.058
&    0.002 &    0.000 & $-$0.003
& $-$0.003 & $-$0.002 &    0.000  \\
$n=2$
& $-$0.029
&    0.000 & 0.000 & 0.000
& $-$0.002 & 0.005 & $-$0.003
&    0.008 & 0.000 & $-$0.002
\end{tabular}
\end{ruledtabular}
}
\label{Table_J}
\end{table*}

\begin{table}[h]
    \centering
     \caption{Positions of the four Mn sublattices (in units of the lattice constant) and their corresponding spin directions in the magnetic ground state. The ground-state spin configuration used throughout this work is adopted from Ref.~\cite{burlet-97}.}
    \begin{ruledtabular}
    \begin{tabular}{ l | c | c | c | c }
        $\text{Mn}$ atom & $\text{Mn}_1$  & $\text{Mn}_2$ & $\text{Mn}_3$  &  $\text{Mn}_4$\\
        \hline
         Position & (0, 0, 0) &  (0.5, 0.5, 0) & (0, 0.5, 0.5) & (0.5, 0, 0.5)\\
         Spin direction & $[111]$ & $[1\bar{1}\bar{1}]$ & $[\bar{1}1\bar{1}]$ & $[\bar{1}\bar{1}1]$
    \end{tabular}
    \end{ruledtabular}
    \label{Table1}
\end{table}

The spin Hamiltonian derived from the DFT calculations is mapped onto a magnon Hamiltonian by applying the Holstein-Primakoff transformation~\cite{Holstein-40}, assuming the noncollinear spin ground state specified in Table~\ref{Table1}~\cite{burlet-97}.
Within the linear spin-wave regime, the quadratic magnon Hamiltonian is diagonalized following Colpa's method~\cite{Colpa-78}. From the resulting eigenstates, we compute both the magnon energy dispersion and the associated magnon spin texture by evaluating the expectation values of the magnon Hamiltonian and spin operators, respectively.

\begin{figure*}
    \centering
    \includegraphics[width=500pt]{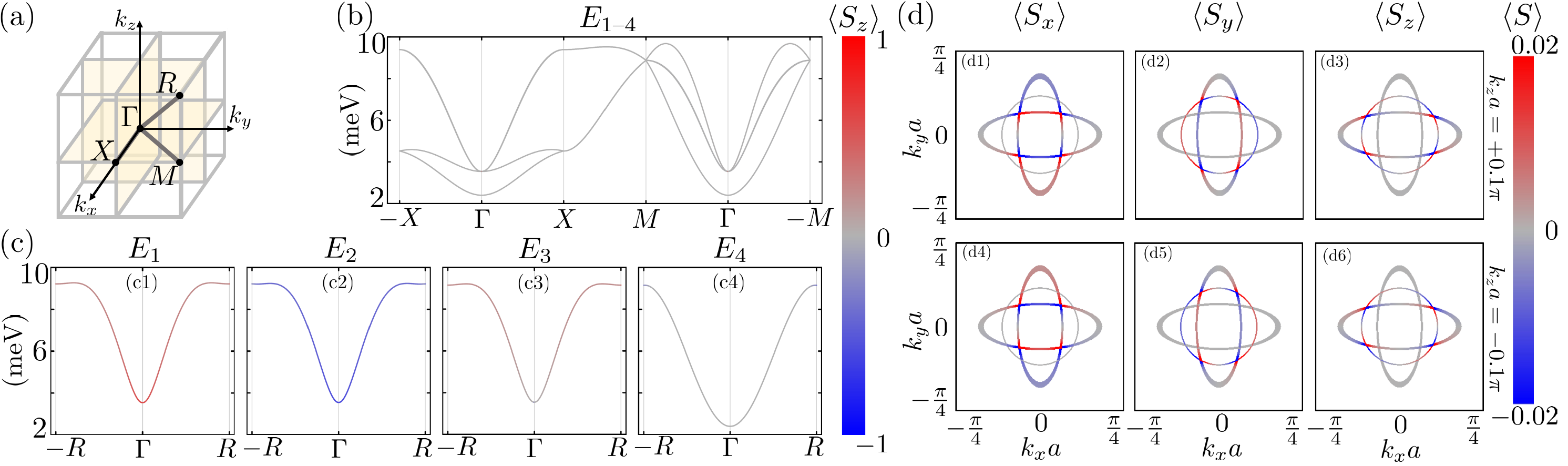}
    \caption{
    (a)~Schematic picture of the symmetric points in the Brillouin zone.
    Magnon energy bands of the $\text{MnTe}_2$ colored by its $z$-component spin texture along the path connecting the high-symmetry points (b)~$(-X)$-$\Gamma$-$X$-$M$-$\Gamma$-$(-M)$ and (c)~$(-R)$-$\Gamma$-$R$.
    (c1-4)~The bands $E_i$s, where $E_i\geq E_j$ for $i>j$, are presented separately to show the spin texture without overlap.
    (d)~Spin textures of the $E_{1\text{-}3}$ bands in the energy window $[3.89\text{ meV},3.91\text{ meV}]$.
    Panels~(d1-3) show the result for the $k_z a=+0.1\pi$ plane, and panels~(d4-6) show $k_z a=-0.1\pi$ plane.
    Panels (d1, d4), (d2, d5), and (d3, d6) display the $x$, $y$, and $z$-spin textures of magnon, respectively.
    }
    \label{Band_and_Texture_Fig}
\end{figure*}

The four magnon bands are shown in Fig.~\ref{Band_and_Texture_Fig}(b) along the high-symmetry path $(-X)$-$\Gamma$-$X$-$M$-$\Gamma$-$(-M)$ in the Brillouin zone, connecting the high-symmetry points $\Gamma$, $X$, $M$ and $X$, which are located at $(0,0,0)$, $(1,0,0)\frac{\pi}{a}$, and $(1,1,0)\frac{\pi}{a}$, respectively [Fig.~\ref{Band_and_Texture_Fig}~(a)].
At the $\Gamma$ point, the three upper magnon bands are degenerate with energies $E_1=E_2=E_3=3.54\text{ meV}$, while the lowest band lies at $E_4=2.40\text{ meV}$.
These theoretical values are in good agreement with the experimentally observed magnetic excitations at $2.5 \text{ meV}$ and $3.6 \text{ meV}$, thereby validating the magnetic coupling constants extracted from the DFT calculations as well as the subsequent spin-wave analysis.
At the $\pm X$ and $\pm M$ points as well as along the $X-M$ line, the magnon bands are doubly degenerate $E_1=E_2\neq E_3=E_4$.
In Fig.~\ref{Band_and_Texture_Fig}(b), the spin-$z$ polarization of the magnon along this path is negligibly small, remaining below 0.01.
On the other hand, Fig.~\ref{Band_and_Texture_Fig}(c) shows a strong spin-$z$ texture along the path $(-R)$-$\Gamma$-$R$, along which the magnon bands are non-degenerate except at the $\Gamma$ point and $\pm R$ points, where the $R$ point is located at $(1,1,1)\frac{\pi}{a}$ [Fig.~\ref{Band_and_Texture_Fig}(a)].

The altermagnetic nature of the magnon bands, i.e., their direction-dependent spin polarization, is clearly seen in Fig.~\ref{Band_and_Texture_Fig}(d), which displays the spin textures of the three upper magnon bands on the $k_z a = \pm0.1\pi$ planes.
Specifically, the spin textures shown in Fig.~\ref{Band_and_Texture_Fig}~(d) satisfy the symmetry relations that $\langle S_i\rangle$ is an even function under $k_i\rightarrow -k_i$ and an odd function under $k_j\rightarrow -k_j$, where $i,j\in\{x,y,z\}$ with $j\neq i$. This plaid-like spin-splitting structure of the magnon bands is identical to that previously identified in the electronic band structure~\cite{Zhu-24}.
The common symmetry properties of the momentum-space spin textures in both magnonic and electronic bands originate from the symmetry of $\text{MnTe}_2$  under the combined operation $[2_{\hat{\boldsymbol{e}}^{i}}|| \{2_{\hat{\boldsymbol{e}}^{i}} |\frac{a}{2}(\hat{\boldsymbol{e}}^{i}+\hat{\boldsymbol{e}}^{j}) \}$~\cite{Corticelli2022, Chen2024}, where $i,j\in\{x,y,z\}$ are chosen in a cyclic way.
For example, for $i=y$ and $j=z$, the ground configuration of $\text{MnTe}_2$ is invariant under the composite operation consisting of a $\pi$ rotation of the spins about the $y$-axis, a $\pi$ rotation of the lattice about the $y$-axis, and a lattice
translation by $(0,0.5a,0.5a)$.
This feature is consistently shown for other planes, i.e., on the $k_x a=\pm 0.1\pi$ and $k_y a=\pm 0.1\pi$ planes.
Our theoretically calculated altermagnetic spin texture of $\text{MnTe}_2$
is fully consistent with our experimental observation of a pronounced $RL/LR$ Raman intensity imbalance, indicating that, for a given magnon propagation direction, one chirality is energetically favored over the other.

\section{Discussion}
Our combined experimental and theoretical results establish $\text{MnTe}_2$ as a compelling realization of noncollinear altermagnetism in a three-dimensional pyrite-type lattice.  The experimentally observed RCD
in the one-magnon excitations provides a direct optical signature of the symmetry-broken state. The most striking feature is its antisymmetric reversal between the Stokes and anti-Stokes channels for the 2.5 meV magnon branch.
In conventional antiferromagnets, magnons typically occur as left- and right-handed partners that are exactly degenerate in energy.  In MnTe$_2$, however, the unconventional crystalline symmetries can lift this degeneracy for both electronic and magnon bands. This energy splitting between chiral partners ensures that, for a given propagation direction, one handedness is energetically favored, leading to the observed polarization-dependent scattering selection rules.

Notably, the chiral anomaly is pronounced in the lower-energy branch, whereas the higher-energy
$T_g$ magnons exhibit no discernible difference between the Stokes and anti-Stokes scattering channels. Linear spin-wave theory predicts a larger spin polarization for the three high-energy magnon bands than for the low-energy magnon band. At first glance, this appears inconsistent with experimental observations, which show a stronger chirality imbalance in the low-energy sector (~2.5 meV) than in the high-energy sector (~3.6 meV). However, the Raman intensity imbalance, although correlated with the spin polarization of the corresponding magnons, is also affected by several additional factors, including the scattering cross-section and the structure of the Raman vertex. Consequently, the finite spin polarization obtained from spin-wave theory should be understood as evidence for the existence of a Raman chirality imbalance, rather than as a quantitative explanation of its magnitude.

Furthermore, the angle-resolved Raman data reveal the reduced rotational symmetry of the magnon Raman response expected for the fourfold symmetry, reflecting the breaking of combined spatial and temporal symmetries in momentum space.
Crucially, our DFT+$U$ and linear spin-wave theory calculations reveal a plaid-like spin-splitting texture in momentum space. This texture is a bosonic analogue to the electronic band splitting previously identified in altermagnets~\cite{Zhu-24}. The characteristic even-odd momentum dependence of the magnon spin polarization is not driven by relativistic spin–orbit coupling or conventional antisymmetric exchange. Instead, it is enforced by composite spin-lattice operations—specifically rotations and fractional translations that connect distinct magnetic sublattices. In particular,
the 90$^{\circ}$ rotation of the opposite spin sublattice provides a natural explanation for the emergence of the twofold symmetry observed in the magnon modes.

More broadly, our findings  demonstrate that noncollinear antiferromagnets can host robust, symmetry-protected chiral magnons. The ability to access these symmetry-imposed spin textures via the table-top and cost-efficient method of polarization- and helicity-resolved Raman spectroscopy provides a powerful route to experimentally classify altermagnets based on their reciprocal-space symmetries rather than solely on static spin configurations. This approach complements neutron scattering and photoemission techniques and is particularly advantageous for materials with small sample volumes or complex magnetic unit cells. Ultimately, the symmetry-driven control of magnon chirality demonstrated here points toward new opportunities for engineering non-reciprocal magnonic functionalities and directional spin transport in altermagnetic spintronics.

\section{\label{sum}Conclusion}
In conclusion, our study identifies the pyrite-structured MnTe$_2$ as a prime realization of a non-collinear altermagnet, where the interplay of complex magnetic sublattices and crystalline symmetries engenders unconventional spin-wave excitations. Through helicity- and angle-resolved Raman spectroscopy, we have observed a definitive circular dichroism and a striking Stokes/anti-Stokes intensity imbalance in the low-energy magnon branch, signaling the presence of magnetic excitations with a well-defined handedness. These experimental findings are supported by DFT+$U$ and linear spin-wave theory calculations, which reveal a characteristic plaid-like spin-splitting texture in the magnon bands. This work  provides a robust experimental framework for exploring the emerging physics of chiral quasiparticles and non-reciprocal transport in high-symmetry compensated non-coplanar magnets.

\section{\label{exp}Experimental Section}

{\it Sample synthesis:} High-quality single crystals of manganese ditelluride (MnTe$_2$) were grown using the chemical vapor transport (CVT) method with iodine as the transport agent. Polycrystalline precursors were first synthesized via a conventional solid-state reaction. Manganese powder and tellurium slugs, both with 3N purity, were mixed in the appropriate stoichiometric ratio and sealed in an evacuated quartz ampoule under high vacuum. The mixture was sintered at 600$^{\circ}$C, with intermittent grinding to ensure phase homogeneity. For crystal growth, approximately 180 mg of iodine was added to the synthesized powder, which was then loaded into a 350 mm-long quartz ampoule and sealed under high vacuum. The ampoule was placed in a two-zone horizontal furnace and maintained under a temperature gradient of 600$^{\circ}$C (source) to 540$^{\circ}$C (sink) for 200 h. Upon completion of the growth process, the furnace was gradually cooled to room temperature at a rate of 2$^{\circ}$C/min. Shiny cubic single crystals with typical dimensions of approximately 1 mm $\times$ 1 mm $\times$ 1 mm were collected from the cold end of the ampoule.\\

{\it Raman spectroscopy:} Temperature- and polarization-resolved Raman scattering experiments were carried out using a diode-pumped continuous-wave laser emitting at 515 nm (Cobolt 05-01 series). A freshly-cleaved sample was directly mounted onto the cold-finger of an open-cycle He-flow cryostat (Oxford) inside an argon-filled glovebox to avoid sample degradation. Thermal contact was between sample and cold-finger was achieved via silver glue (Ted Pella, Inc.). The laser was focused onto the sample with a spot of about 2 $\mu$m in diameter and a laser power of less than 150 $\mu$W to reduce local laser heating effects. Based on a comparison between Stokes- and anti-Stokes intensities measured at various incident laser powers, we estimate a local heating effect of 5 K. All temperatures have been corrected accordingly. The polarization was selected using super-achromatic $\lambda/2$- and $\lambda/4$-waveplates (Thorlabs). The laser lines were discriminated with volume Bragg grating sets (OptiGrate) which allow a low-energy spectral cut-off at 6 cm$^{-1}$ or less. The inelastically scattered light was dispersed through a single-stage spectrometer (Princeton Instruments HRS-750, 1800 gr/mm) and recorded by a nitrogen-cooled charge-coupled device (PyLoN eXcelon).

{\it DFT calculations:}
DFT+$U$ calculations were carried out using the projector augmented-wave (PAW) method as implemented in the Vienna \emph{Ab initio} Simulation Package (VASP)~\cite{vasp1,vasp2}. Atomic positions were optimized until the residual forces became smaller than 1~meV/${\AA}$ while fixing the experimental lattice constant to 6.90~${\AA}$. An energy cutoff of 500~eV and a $10\times10\times10$ $\Gamma$-centered $k$-mesh were used.
To obtain the full-band Hamiltonian for magnetic force calculations, we additionally performed DFT+$U$ calculations using the OpenMX package~\cite{openmx_package,han2006n}, which employs a localized pseudoatomic orbital basis set. We chose the basis sets: $s3p2d2f1$ for Mn and $s3p3d2f1$ for Te. In these calculations, a $9\times9\times9$ $k$-mesh and an energy cutoff of 1000~Ry were used.

For the exchange--correlation functional, we adopted the generalized gradient approximation (GGA) in the Perdew–Burke–Ernzerhof (PBE) form~\cite{pbe}. For better comparison with Refs.~\cite{Zhu-24, xu2018mnte2}, we mainly present the results of DFT+$U$ calculations with $U_{\rm eff}=4.2$~eV for Mn-$3d$ electrons, as first suggested by Dudarev and co-workers~\cite{dudarev_U,han2006n}. Varying $U_{\rm eff}$ in the range of 1.0--5.0~eV, the ratio $|\vec{D}_1|/J_1$ changes by less than 20.3~\%, indicating the robustness of our results against reasonable variations of $U_{\rm eff}$.

In addition, we performed calculations within the charge-only DFT implementation of the Liechtenstein formalism (say, cFLL$+U+J$)~\cite{liechtenstein1995density, ryee2018effect, ryee2018comparative} with $U = 5.0$~eV and $J = 0.8$~eV. The most pronounced difference in the electronic structure is a downward shift of the Mn-$3d$ levels by approximately 0.3~eV in cFLL$+U+J$, accompanied by modified band dispersions. These changes result in quantitative differences in the magnetic coupling constants: $J_1 = 1.459$~meV and $|\vec{D}_1| = 0.107$~meV, corresponding to an approximately 21\% change in the ratio $|\vec{D}_1|/J_1$, compared to the results obtained using the Dudarev's functional. Magnetic interaction parameters were computed based on the magnetic force linear-response theory~\cite{liechtenstein1987} using the \texttt{Jx} code~\cite{yoon2018reliability,yoon2020jx}.

\begin{acknowledgments}
D.W. was supported by the Institute of Applied Physics of Seoul National University, by the ITRC Program through the IITP, and by the Global Research Development Center (GRDC) Cooperative Hub Program through the NRF of Korea, funded by the (MSIT) (Grants No. RS-2024-00437191 and RS-2023-00258359). K.-Y.C. was supported by the National Research Foundation (NRF) of Korea (Grants No. 2020R1A2C3012367 and No. 2020R1A5A1016518).
D.A. and S.K.K. were supported by Brain Pool Plus Program through the National Research Foundation of Korea funded by the Ministry of Science and ICT (2020H1D3A2A03099291) and Basic Science Research Program through the National Research Foundation of Korea (NRF) funded by the Ministry of Education (2019R1A6A1A10073887).
J.C. and M.J.H. were supported by the National Research Foundation of Korea (NRF) grant funded by the Korea government (MSIT) (Grant Nos. RS-2025-02243032 and RS-2025-00559042). R.S. acknowledges the financial support provided by the Ministry of Science and Technology in Taiwan under Project Numbers NSTC- 114-2124-M-001-009, NSTC-113-2112-M-001-045-MY3, Financial support from the Center of Atomic Initiative for New Materials (AI-Mat), National Taiwan University (Project No. 113L900801) and Academia Sinica for the budget of AS-iMATE-115-11.
\end{acknowledgments}

\section*{Conflict of Interest}
The authors declare no conflict of interest.

\section*{Author Contributions}

\section*{Data Availability Statement}
The data that support the findings of this study are available from the corresponding
author upon reasonable request.

\section*{Keywords}
altermagnet, chiral magnon, Raman spectroscopy, DFT calculations

\end{document}